\documentclass[aps,prd,showkeys,nofootinbib,amsmath,amssymb,superscriptaddress,twocolumn]{revtex4-2}
\usepackage{amsfonts}
\usepackage{amsthm}
\usepackage[title]{appendix}
\usepackage{booktabs}
\usepackage{dsfont}
\usepackage{enumerate}
\usepackage{float}
\usepackage[T1]{fontenc}
\usepackage{framed}
\usepackage[left=2cm,right=2cm,top=2cm,bottom=2cm]{geometry}
\usepackage{graphicx}
\usepackage{hyperref}
\usepackage[utf8]{inputenc}
\usepackage{multirow}
\usepackage{scalerel}[2014/03/10]
\usepackage{slashed}
\usepackage{tabularx}
\usepackage[normalem]{ulem}
\usepackage{stackengine}
\usepackage{titlesec}

\usepackage{color}

\newcommand{\diff}{\mathrm{d}}

\newcommand{\ph}{\ensuremath{\hat{p}}}
\newcommand{\qh}{\ensuremath{\hat{q}}}
\newcommand{\Ph}{\ensuremath{\hat{P}}}
\newcommand{\Qh}{\ensuremath{\hat{Q}}}

\newcolumntype{R}{>{\raggedleft\arraybackslash}X}
\newcolumntype{L}{>{\raggedright\arraybackslash}X}

\begin{document}

\title{Space and time transformations with a minimal length}

\author{Pasquale Bosso}\email[]{bosso.pasquale@gmail.com}
\affiliation{University of Lethbridge, Department of Physics and Astronomy, 4401 University Drive, Lethbridge, Alberta, Canada, T1K 3M4}

\date{\today}

\begin{abstract}
  Phenomenological studies of quantum gravity have proposed a modification of the commutator between position and momentum in quantum mechanics so to introduce a minimal uncertainty in position in quantum mechanics.
  Such a minimal uncertainty and the consequent minimal measurable length have important consequences on the dynamics of quantum systems.
  In the present work, we show that such consequences go beyond dynamics, reaching the definition of quantities such as energy, momentum, and the same Hamiltonian.
  Furthermore, since the Hamiltonian, defined as the generator of time evolution, results to be bounded, a minimal length implies a minimal time.
\end{abstract}

\maketitle

Models of quantum mechanics with a minimal length are motivated by phenomenological studies of quantum gravity since several candidate theories have proposed the existence, under several perspectives and in different shapes, of a minimal accessible length 
\cite{Mead:1964zz,Gross:1987kza,Gross:1987ar,Amati:1987wq,Amati:1988tn,Konishi:1989wk,Garay:1994en,Adler:1999bu,Scardigli:1999jh}.
One of the most common approaches for such phenomenological studies consists of a modification of the Heisenberg algebra describing a modified uncertainty relation between position and momentum, generically called Generalized Uncertainty Principle (GUP)
\cite{Maggiore:1993kv,Kempf:1994su,Das:2008kaa,Ali:2011fa,Lake:2018zeg,Bosso:2020aqm,Petruzziello:2020wkd,Wagner:2021bqz,Bosso:2021koi,Gomes:2022akh}.
Such a modified algebra is then used to study several aspects of quantum systems, such as energy spectra and dynamical evolution, which are then compared with observations or are further studied to propose experiments able to observe the effects of a minimal length
\cite{Pikovski:2011zk,Bawaj:2014cda,Bosso:2018ckz,Bushev:2019zvw,Bosso:2016ycv}.
The objective of the present work is to study the premises of such studies.
Specifically, we intend to investigate quantities, such as energy, which may be affected by a minimal length and how they are affected under the assumptions of Galilean Relativity.
Furthermore, special focus will be here given to time evolution and the corresponding generator.
We anticipate that a minimal length decouples energy from the time evolution, thus requiring a new and more accurate analysis of the energy eigenvalue equation in the context of GUP.
The hypothesis of Galilean Relativity will be crucial, as it determines the relations of the Hamiltonian with the other generators.

This work is structured as follows: 
in Section \ref{sec:1d}, we study the algebra of the generators for Galilean transformations in one spatial dimension, describing their physical interpretation.
In Section \ref{ssec:energy}, we derive a form for the kinetic and potential energies, thus showing that the total energy is distinct from the Hamiltonian of a system.
In Section \ref{ssec:2degree}, we specialize the results for a specific deformation, proposing three examples:
a particle in a square box, the case of a square barrier (Section \ref{sssec:box_barrier}), and a harmonic oscillator (Section \ref{sssec:HO}).
In Section \ref{sec:3d}, we repeat the analysis of the algebra of the generators in three spatial dimensions.
We conclude in Section \ref{sec:conclusions} with a summary of the work and remarks.

\section{One dimension}
\label{sec:1d}

Let us start with a generic one-dimensional model of the form
\begin{equation}
  [\hat{q},\hat{p}]
  = i \hbar f(\hat{p}).
  \label{eqn:GUP_1}
\end{equation}
Details for this model can be found in \cite{Bosso:2020aqm,Bosso:2021koi}.
Here, we want to study the transformation properties of a system described by such a model and the relations between physical quantities such as position, momentum, and energy, and the generators of transformations.
Let us thus start with a brief review of non-relativistic space and time symmetry transformations in quantum mechanics, as described in \cite{ballentine2014quantum}.
Specifically, we will consider the properties of continuous transformations, such as displacement, time translations, and Galilean boosts.
In the next section, in which we will consider extensions to three dimensions, we will also include rotations.
As usual, such transformations will be described in terms of unitary, linear operators acting on state vectors.
That is, given any such transformations described in terms of an operator $\hat{U}$, we require the following transformation rules for a state vector $|\Psi\rangle$ and an operator $\hat{A}$
\begin{align}
  |\Psi\rangle \to & |\Psi'\rangle = \hat{U} |\Psi\rangle,&
  \hat{A} \to & \hat{A}' = \hat{U} \hat{A} \hat{U}^{-1}.
  \label{eqns:transformations}
\end{align}
The transformed state and operator are related to the original ones in such a way that, considering an eigenstate $|\Phi_n\rangle$ of the operator $\hat{A}$ with eigenvalue $a_n$, the transformed state $|\Phi_n'\rangle$ is an eigenstate of the transformed operator $\hat{A}'$ with the same eigenvalue.
Furthermore, the probability of equivalent events is invariant under such transformations, that is $|\langle \Phi_n | \Psi \rangle|^2 = |\langle \Phi_n' | \Psi' \rangle|^2$.

Since the unitary operator $\hat{U}$ will be associated with continuous transformations, we can in principle identify some Hermitian operators acting as generators for the corresponding transformations.
In other words, given a transformation described by the unitary operator $\hat{U}$, such an operator can be written as
\begin{equation}
  \hat{U} = e^{i \hat{G} x},
  \label{eqn:generator}
\end{equation}
where $\hat{G}$ is the generator of the transformation, while $x$ parametrizes the transformation.

As mentioned above, here we will consider displacements by a distance $d$, time translations by an interval $\tau$, and changes of reference frame with relative velocity $v$.
Thus, given a point at a position $q$ and an instant $t$, a generic transformation changes the position and the instant of time as
\begin{align}
  q \to q' &= q + d + v t,&
  t \to t' &= t + \tau.
  \label{eqn:transformations_1D}
\end{align}
Associated with such transformations, we introduce the following unitary operators and corresponding generators
\begin{align}
&& \centering \text{Unitary}&\text{ operators} & \centering \text{Genera}&\text{tors}\nonumber\\
  &\parbox{7em}{Time evolution}&
  \hat{U}_T(\tau) &= e^{i \hat{H} \tau / \hbar}&
  \hat{H}
  \label{eqn:time_evol_gen}\\[1em]
  &\parbox{7em}{Spatial translation}&
  \hat{U}_S(d) &= e^{- i \hat{P} d / \hbar}&
  \hat{P}
  \label{eqn:space_trans_gen}\\[1em]
  &\parbox{7em}{Change of reference frame}&
  \hat{U}_F(v) &= e^{i \Qh v m / \hbar}&
  \Qh
  \label{eqn:change_ref_frame_gen}
\end{align}
where $m$ is a constant with units of a mass.

By considering suitable combinations of transformations, it is possible to find the following set of commutation relations for a system symmetric under all Galilean transformations \cite{ballentine2014quantum}
\begin{align}
  [\Qh,\Ph] &= i \hbar,&
  [\Qh,\hat{H}] &= i \Ph \hbar / m,&
  [\Ph,\hat{H}] &= 0.
  \label{eqns:commutators_1d}
\end{align}
However, it is important to notice that, up to now, we have not assigned any physical meaning to the generators, besides associating them with the corresponding transformations.
In the standard theory, the three operators $\Qh$, $\Ph$, and $\hat{H}$ are identified with the position operator, momentum operator, and the Hamiltonian operator.
We will soon see that, although such operators can still be associated with such physical quantities, in general, identification is not possible when GUP is accounted for.

Consider first the position operator $\hat{q}$.
In models of quantum mechanics with a minimal measurable length, the eigenstates of such an operator are not physical \cite{Kempf:1994su}.
That is, we cannot identify a hypothetical state $|q\rangle$ such that $\hat{q} |q\rangle = q |q\rangle$.
Nonetheless, any state $|\Psi\rangle$ is characterized by a specific expectation value of the position, \emph{i.e.} $\langle \Psi | \hat{q} | \Psi \rangle = \tilde{q}$.
Therefore, we expect that translating the state via the generator $\Ph$ would result in a new state $|\Psi'\rangle$ characterized by a different expectation value of the position.
Specifically, suppose the state is translate by a distance $d$.
Then,
\begin{equation}
  \langle \Psi' | \hat{q} | \Psi' \rangle = \tilde{q} + d, \qquad \text{with } |\Psi'\rangle = e^{-i \Ph d / \hbar} |\Psi\rangle.
\end{equation}
The same relation is trivially equivalent to the expectation value on the original state $|\Psi\rangle$ of the position operator $\qh'$ translated by $-d$.
Now, since
\begin{equation}
  \langle \Psi' | \hat{q} | \Psi' \rangle
  = \langle \Psi | \hat{q}' | \Psi \rangle
  = \tilde{q} + d
  = \langle \Psi | \hat{q} | \Psi \rangle + d,
\end{equation}
for any state $|\Psi\rangle$, we need to have $\hat{q}' = \hat{q} + d$.
Thus, considering the transformation \eqref{eqns:transformations} for an infinitesimal displacement $d$, we obtain
\begin{equation}
  \hat{q}' = e^{i \Ph d / \hbar} \hat{q} e^{-i \Ph d / \hbar} = \hat{q} + d \quad \Rightarrow \quad [\hat{q},\Ph] = i \hbar.
\end{equation}
Comparing this last commutator with the first of \eqref{eqns:commutators_1d}, we notice that $\hat{Q} = \hat{q}$
\footnote{A more detailed demonstration of this equality is given in \cite{Jordan} and in the references therein, as well as in \cite{ballentine2014quantum}.}.
In other words, the position operator $\hat{q}$ corresponds to the generator of changes of reference frame.
Similarly, we notice that the commutator between the position operator and the generator of spatial translations $\Ph$ does not correspond to \eqref{eqn:GUP_1}.
Therefore, we can conclude that, when a GUP model is involved, the momentum operator $\hat{p}$ and the generator of translations $\Ph$ are two distinct operators.
However, it is easy to see that they are one the function of the other, $\Ph = P(\ph)$.
Specifically, we can write \cite{Bosso:2021koi}
\begin{equation}
  [\qh,\ph]
  = i \hbar \widehat{\frac{\diff p}{\diff P}}
  = i \hbar f(\ph).
  \label{eqn:f_as_derivative}
\end{equation}
This result is consistent with what is found in \cite{Bosso:2020aqm}.
It is worth notice that, in a classical theory, the operator $\Ph$ would be associated with the momentum $P$ conjugate to the physical position $q$ \cite{Bosso:2018uus,Bosso:2022vlz}.
We will thus refer to the operator $\Ph$ with the expression ``canonical momentum'' too.

As for the generator of time translations $\hat{H}$, considering a generic time-dependent vector state, we have the following equation, describing the time evolution of the state
\begin{equation}
  i \hbar \frac{\diff}{\diff t} |\Psi(t)\rangle = \hat{H} |\Psi(t)\rangle.
  \label{eqn:time_evo_state}
\end{equation}
Furthermore, by \eqref{eqns:commutators_1d}, up to a constant we get
\begin{equation}
  \hat{H} = \frac{\Ph^2}{2 m}.
  \label{eqn:time_generator_free}
\end{equation}
Although this expression presents the same form of a free Hamiltonian in quantum mechanics, it is worth emphasizing that $\Ph$ is not the physical momentum operator, as well as $\hat{H}$ cannot be associated with the energy of a system.

When interactions are considered, the generator $\hat{H}$ has to be changed to include the effects of the interactions on the time evolution of states and operators.
However, Eq. \eqref{eqn:time_evo_state} is still valid given the role of $\hat{H}$ as generator of time evolution.
Adopting the Schr\"odinger picture and using \eqref{eqn:time_evo_state}, we then find that the velocity operator $\hat{v}$, the operator defined such that
\begin{equation}
  \langle \hat{v} \rangle
  = \frac{\diff}{\diff t} \langle \qh \rangle,
  \label{eqn:vel_q}
\end{equation}
is given by
\begin{equation}
  \hat{v}
  = \frac{1}{i \hbar } [\qh,\hat{H}].
\end{equation}
Considering $\hat{q}$ as generator of changes of reference frame, its commutator with the velocity operator has to be given by
\begin{equation}
  [\qh,\hat{v}] = i \hbar / m.
\end{equation}
In the free case this implies
\begin{equation}
  \hat{v} = \frac{\Ph}{m}.
  \label{eqn:vel_gen}
\end{equation}
Due to \eqref{eqn:time_generator_free} and \eqref{eqn:vel_gen}, we can associate to $m$ the meaning of mass of the system.
Furthermore, it is worth noticing that the product of mass and velocity produces the generator of translations $\Ph$ rather than the physical momentum operator $\ph$.
However, in the presence of interactions, as $\hat{v} - \Ph/m$ commutes with $\qh$, such a difference may be a function of $\qh$, that is
\begin{equation}
  \hat{v} = \frac{\Ph - C(\qh)}{m}.
\end{equation}
For the same reason, since the difference between $\hat{H}$ and $\Ph^2/2m$ commutes with $\qh$, we can write
\begin{equation}
  \hat{H}
  = \frac{[\Ph - C(\qh)]^2}{2m} + U(\qh).
  \label{eqn:gen_time_evol_gen}
\end{equation}
This is the most general form of the Hamiltonian of a spinless system compatible with Galilei invariance and GUP in one dimension.
We notice again that the form of the generator $\hat{H}$ is extremely similar to that of the Hamiltonian in ordinary quantum mechanics.
Nonetheless, we can see that, in models characterized by a modified position-momentum commutator, it cannot be directly associated with the energy of a system.

\subsection{Energy}
\label{ssec:energy}

To identify the physical quantity energy, we will start from classical mechanics.
Furthermore, for the moment we will consider the case of a vanishing function $C$ in Eq. \eqref{eqn:gen_time_evol_gen}.

Consider a force as the physical quantity measured by a dynamometer.
Furthermore, we define the physical momentum $p$ of an object acted upon by a system of forces as the quantity such that
\begin{equation}
  F = \dot{p},
  \label{eqn:force}
\end{equation}
with $F$ the net force acting on the object and where a dot signifies a time derivative.
Given the relation between the physical momentum $p$ and the canonical momentum $P$ in \eqref{eqn:f_as_derivative}, we can write
\begin{equation}
  F = f(p) \dot{P}.
  \label{eqn:force_P}
\end{equation}
Since the physical position $q$ and the conjugate momentum $P$ form a set of canonical variables, as implied by \eqref{eqns:commutators_1d} and by several previous works \cite{Bosso:2018uus,Bosso:2021koi,Bosso:2020aqm,Wagner:2021bqz,Bosso:2022vlz}, we can find the following Hamilton equations based on an Hamiltonian of the form \eqref{eqn:gen_time_evol_gen}
\begin{align}
  \dot{q} &= \frac{\partial H}{\partial P} = \frac{P}{m},
  \label{eqn:Hamilton_q}\\
  \dot{P} &= - \frac{\partial H}{\partial q} = - \frac{\partial U}{\partial q}.
  \label{eqn:Hamilton_P}
\end{align}
Now, let us define the work done by the net force $F$ as the line integral of the force along some path $\gamma$
\begin{equation}
  W = \int_\gamma \dot{p} ~ \diff q.
\end{equation}
In classical mechanics, the work done by a force acting on a system corresponds to the change in kinetic energy of the same system.
Thus, considering an infinitesimal work, we can write
\begin{equation}
  \dot{p} ~ \diff q
  = \dot{q} ~ \diff p.
\end{equation}
where $\dot{q}$ is undestood as the velocity of the system.
Thus, the corresponding change in the kinetic energy $T$ is
\begin{equation}
  \diff T
  = \dot{q} ~ \diff p.
  \label{eqn:free_kinetic_energy}
\end{equation}
Since from the correspondence principle applied to \eqref{eqn:vel_gen} we find
\begin{equation}
  \dot{q} = \frac{P}{m},
  \label{eqn:free_velocity}
\end{equation}
we get the following equivalent expressions for the kinetic energy
\begin{equation}
  \begin{array}{rl}
    T &= \displaystyle \int \dot{q} ~ \diff p\\[1em]
    &= \displaystyle \frac{1}{m} \int P(p) ~ \diff p\\[1em]
    &= \displaystyle \frac{1}{m} \int (f\circ p)(P) P ~ \diff P,
  \end{array}
  \label{eqn:kinetic_energy}
\end{equation}
with the arbitrary constant being fixed so that $T=0$ for $p=0$ or $P=0$.
Furthermore, it is worth noticing that, since $P$, and therefore $\dot{q}$, does not need to be an odd function of the physical momentum $p$, depending on the specific model, we may find left- and right-moving particles characterized by the same physical momentum but different kinetic energies, as we will see below with an example.
Finally, we observe that the kinetic energy has the correct limit for small values of the physical and canonical momenta.
Specifically, we get
\begin{equation}
  T \begin{aligned}[t]
    &= \frac{p^2}{2m} + \mathcal{O}(p^3)\\
    &= \frac{P^2}{2m} + \mathcal{O}(P^3).
  \end{aligned}
\end{equation}
However, it is important to notice that, in general, the kinetic energy cannot be associated with $\frac{p^2}{2m}$ or $\frac{P^2}{2m}$.

The expression we got for the kinetic energy $T$ corresponds to the energy of a free system up to a constant.
With the introduction of a potential-like term in \eqref{eqn:gen_time_evol_gen},
we can find an expression for the potetial energy.
Specifically, we will define as potential energy related to the function $U(q)$ in the Hamiltonian the work, with opposite sign, done by the corresponding force.
From \eqref{eqn:Hamilton_P}, we get that the force related with the term $U(q)$ is
\begin{equation}
  F 
  = \dot{p}
  = - f(p) \frac{\diff U}{\diff q}.
\end{equation}
It is worth observing that such a force depends on the momentum $p$.
Since in general the momentum $p$ depends implicitly on the position because of the forces acting on the system, we can write for the potential energy $U_E$
\begin{equation}
  \diff U_E 
  = - \dot{p} ~ \diff q
  = f(p(q)) \frac{\diff U}{\diff q} ~ \diff q.
  \label{eqn:potential_energy}
\end{equation}
Thus, to find the potential energy of a system, one first needs to find how the physical momentum $p$ depends on the position $q$.
This can be done by first solving the equations of motion \eqref{eqn:Hamilton_q} and \eqref{eqn:Hamilton_P}, as we will see in an example below.
Finally, the energy can be found as the sum of the kinetic and potential terms, from \eqref{eqn:free_kinetic_energy} and \eqref{eqn:potential_energy},
\begin{equation}
  E = T + U_E.
  \label{eqn:total_energy}
\end{equation}
It is worth observing that, given the definitions of kinetic and potential energies, the sum between the two is a constant.
This statement can be made more precise by noting that the Poisson bracket between the energy in \eqref{eqn:total_energy} and the classical counterpart of the Hamiltonian in \eqref{eqn:gen_time_evol_gen} vanishes.
Specifically,
\begin{equation}
  \{E,H\}
  = \frac{\diff U_E}{\diff q} \frac{P}{m} - \frac{\diff T}{\diff P} \frac{\diff U}{\diff q}
  = 0,
\end{equation}
where the Poisson bracket has been computed in terms of the conjugate variables $q$ and $P$ \cite{Bosso:2018uus,Bosso:2022vlz}.

Now, we are going to promote the classical quantity $E$ to an operator.
A first result we can obtain is that the dispersion relation for non-relativistic particles is preserved.
To see this, consider a free system, for which we have \eqref{eqn:time_generator_free}.
An eigenstate of the generator of translations is also an eigenstate of the generator of time evolution.
In (quasi-)position representation, the spatial dependence is given by a phase characterized by a wave number $k = \tilde{P}/\hbar$, with $\tilde{P}$ the eigenvalue of $\Ph$ \cite{Bosso:2020aqm}.
Furthermore, the time dependence, by \eqref{eqn:time_evo_state}, is given by a phase characterized by an eigenvalue of $\hat{H}$, $\hbar \omega$.
Applying the operator $\hat{H}$ to such a state, we obtain
\begin{equation}
  \hat{H} |\Psi(t)\rangle
  = \frac{\hbar^2 k^2}{2 m} |\Psi(t)\rangle.
\end{equation}
Thus, the relation between frequency and wave number is given by
\begin{equation}
  \omega = \frac{\hbar k^2}{2 m},
\end{equation}
corresponding to the ordinary dispersion relation of non-relativistic free particles.

We conclude this section by noticing that, 
for any specific choice of the model, summarized in the choice of the function $f$ in \eqref{eqn:GUP_1} and relating the momentum operator $\ph$ and the generator $\Ph$, energy can be written as a function of the physical momentum and position, $\hat{p}$ and $\qh$.
However, it is worth highlighting again that in general the kinetic term $\hat{T} \neq \ph^2/2m$.
Such a feature has drastic consequences on the energy spectra of all physical systems, as we will see with some examples below.

\subsection{Second degree model}
\label{ssec:2degree}

\begin{figure}[t]
  \centering
  \includegraphics[width=\columnwidth]{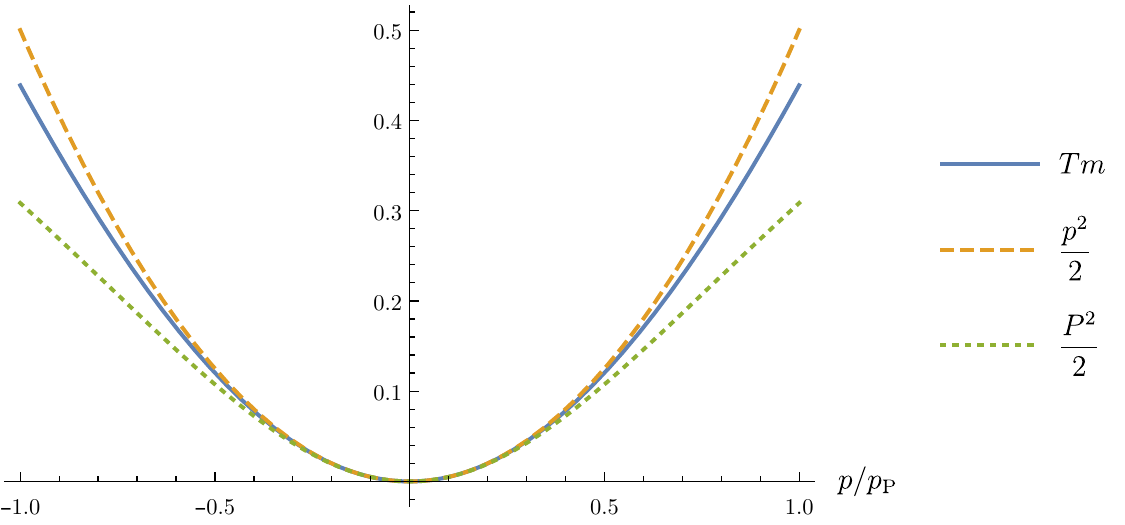}
  \caption{Comparison between the kinetic energy $T$ (solid, blue line), the quantity $\frac{p^2}{2m}$ (dashed, orange line), and $\frac{P^2}{2m}$ (dotted, green line).
  The momentum on the horizontal axis is presented in Planck units, while the three quantities above are presented multiplied by the mass $m$ of the system.
Here, we used $\delta_0 = 0$, $\epsilon_0 = 1$ \cite{Kempf:1994su}.
The vertical axis is given in SI units.}
  \label{fig:kinetic_KMM}
\end{figure}

\begin{figure}[t]
  \centering
  \includegraphics[width=\columnwidth]{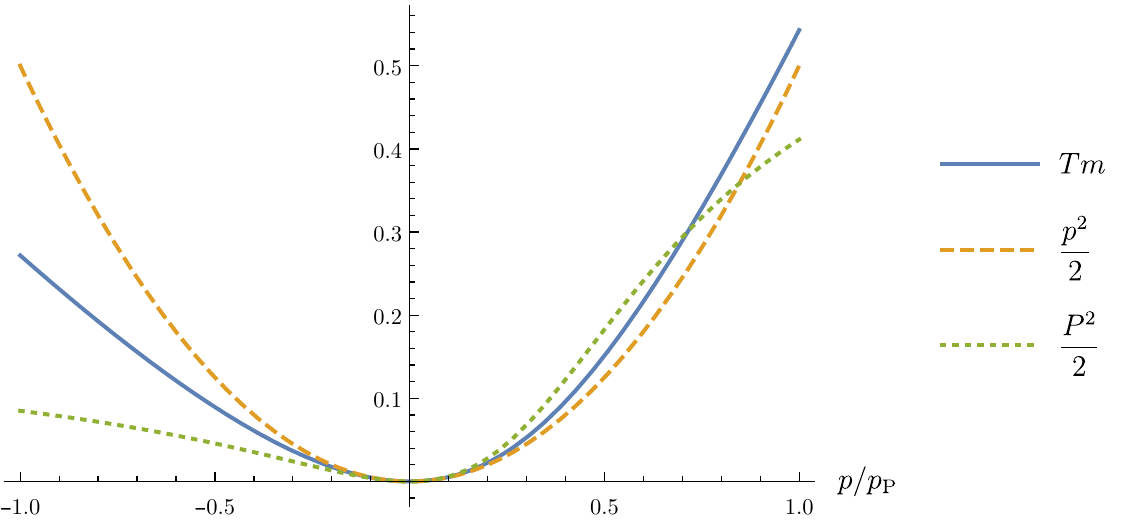}
  \caption{Similar to Figure \ref{fig:kinetic_KMM}.
    Here, we used \mbox{$\delta_0 = 1$, $\epsilon_0 = 3$} \cite{Ali:2011fa}.
It is worth observing that left- and right-moving particles characterized by the same physical momentum $p$ have different kinetic energies $T$ and different values for $\frac{P^2}{2m}$.}
  \label{fig:kinetic_ADV}
\end{figure}

As a specific case, let us consider a model characterized by the following modification function \cite{Bosso:2020aqm}
\begin{equation}
  f(\ph)
  = 1 - 2 \delta \ph + (\delta^2 + \epsilon) \ph^2,
  \label{eqn:deformation_2degree}
\end{equation}
with
\begin{align}
  \delta &= \frac{\delta_0}{p_{\text{P}}},&
  \epsilon &= \frac{\epsilon_0}{p_{\text{P}}^2},
\end{align}
$p_{\text{P}}$ the Planck momentum, $\delta_0$ and $\epsilon_0$ two dimensionless parameters.
With this model, we find the following relations between $P$ and $p$
\begin{align}
  P(p)
  &= \frac{1}{\sqrt{\epsilon}} \arctan \left[ \frac{-\delta + (\delta^2 + \epsilon) p}{\sqrt{\epsilon}} \right] \nonumber \\
  & \qquad + \frac{1}{\sqrt{\epsilon}} \arctan \left( \frac{\delta}{\sqrt{\epsilon}} \right),\\
  p(P)
  &= \frac{\tan \left(\sqrt{\epsilon} P\right)}{\delta  \tan \left(\sqrt{\epsilon} P\right)+ \sqrt{\epsilon }}.
  \label{eqn:pofP_2degree}
\end{align}
We then obtain the following form for the kinetic energy
\begin{align}
  \hat{T} &=
  \begin{aligned}[t]
    & \frac{P(\ph)}{m(\delta^2 + \epsilon)} [\ph (\delta^2+\epsilon) - \delta]\\
    & - \frac{\log \left[1 - 2 \delta \ph + (\delta^2 + \epsilon) \ph^2\right]}{2m(\delta^2 + \epsilon)}
  \end{aligned}\\
  &= \begin{aligned}[t]
    & \frac{\sqrt{\epsilon} \Ph}{m (\delta^2 + \epsilon)} \frac{\sqrt{\epsilon} \sin(\sqrt{\epsilon} \Ph) - \delta \cos(\sqrt{\epsilon} \Ph)}{\delta \sin(\sqrt{\epsilon} \Ph) + \sqrt{\epsilon} \cos(\sqrt{\epsilon} \Ph)} \\
    & + \frac{1}{m (\delta^2 + \epsilon)} \log \left[\frac{\delta}{\sqrt{\epsilon}} \sin (\sqrt{\epsilon} \Ph) + \cos (\sqrt{\epsilon} \Ph)\right].
  \end{aligned}
  \label{eqn:ke_2degree}
\end{align}
We then see that such expressions are different from both $\frac{\ph^2}{2m}$ and $\frac{\Ph^2}{2m}$, as shown in Figures \ref{fig:kinetic_KMM} and \ref{fig:kinetic_ADV} for two specific models.
Furthermore, it is worth noticing that, as $\Ph$ is bounded for this model \cite{Bosso:2020aqm}, free particles are characterized by a frequency that is bounded from above.
In other words, with a maximum wave number, corresponding to a minimum length, comes a maximum frequency, corresponding to a minimum time.

We can now use these results to study some exemplificative physical systems.

\subsubsection{Particle in a box and square potential barrier}
\label{sssec:box_barrier}

Such systems have been studied in \cite{Bosso:2020aqm} by first looking for momentum eigenstates.
Those results still hold, with the only difference of a modified energy spectrum for the particle in a box and modified resonance energies in the transition through the barrier.

\subsubsection{Harmonic oscillator}
\label{sssec:HO}

In this case, the generator $\hat{H}$ reads
\begin{equation}
  \hat{H}
  = \frac{\Ph^2}{2m} + \frac{1}{2} m \omega^2 \qh^2,
  \label{eqn:Hamiltonian_HO}
\end{equation}
where we have assumed that the potential term retains the usual dependence on the position operator.
Because of the first of \eqref{eqns:commutators_1d}, we can define an operator $\hat{a}$ given by
\begin{equation}
  \hat{a} = \sqrt{\frac{m \omega}{2 \hbar}} \left( \qh + i \frac{\Ph}{m \omega} \right),
\end{equation}
and write the generator of time evolution as
\begin{equation}
  \hat{H} = \hbar \omega \left( \hat{a}^\dagger \hat{a} + \frac{1}{2} \right).
\end{equation}
We thus obtain the usual stationary states for a harmonic oscillator when expressed in terms of the physical position $\qh$ and the canonical momentum $\Ph$.

As mentioned above, the canonical momentum $P$ may present a bounded spectrum \cite{Bosso:2020aqm}.
We then expect the Hamiltonian to have a finite, discrete spectrum, with an upper bound determined by the bound on the operator $\Ph$.
Furthermore, since the left and right bound of $\Ph$ may be different, as in the case $\delta_0 = 1$ and $\epsilon_0 = 3$, the bound on the Hamiltonian depends on the smallest of the two bounds of $\Ph$.
To see why the Hamiltonian \eqref{eqn:Hamiltonian_HO} has to be bounded and to find the maximum value of the quantum number corresponding to such a bound, first notice that the expectation value of $\Ph$ on any stationary state vanishes.
As for $\Ph^2$, we have
\begin{equation}
  \langle n | \Ph^2 | n \rangle
  = p_0^2 (2n+1)
  \leq \frac{\arctan^2\left(\frac{\sqrt{\epsilon }}{\delta }\right)}{\epsilon },
\end{equation}
where we introduced the characteristic momentum \mbox{$p_0 = \sqrt{\frac{\hbar \omega m}{2}}$}.
We then get the following expression for the maximum value of the quantum number $n$
\begin{equation}
  n_{\text{max}}
  = \left\lfloor\frac{\arctan^2 \left(\frac{\sqrt{\epsilon}}{\delta}\right)}{4 p_0^2 \epsilon } - \frac{1}{2}\right\rfloor.
\end{equation}
Notice that no physical stationary state is allowed when
\begin{equation}
  p_0
  \leq \frac{\arctan\left(\frac{\sqrt{\epsilon}}{|\delta|}\right)}{\sqrt{2 \epsilon }}.
\end{equation}
Furthermore, the stationary states $|n\rangle$ are characterized by a maximum phase frequency depending on the mass $m$
\begin{equation}
  \Omega
  \leq \omega \left(n_\text{max} + \frac{1}{2}\right) 
  = \frac{\arctan^2 \left(\frac{\sqrt{\epsilon}}{\delta}\right)}{2 \hbar m \epsilon}.
\end{equation}
Such a maximum frequency can be understood as the limit for time intervals over which changes can be detected.
That is, it sets a minimal uncertainty in time.

We can easily find the stationary states in the $P$-representation, that is, in the representation in which $\Ph$ acts as multiplication by $P$.
In such a representation, the position operator is given by \cite{Bosso:2020aqm}
\begin{equation}
  \qh = i \hbar \frac{\diff}{\diff P}.
\end{equation}
Thus, stationary states are expressed in terms of Hermite polynomials as in ordinary quantum mechanics, namely
\begin{equation}
  \phi_n(P)
  = \langle P | n \rangle
  = \mathcal{N} 2^{-\frac{n}{2}} e^{-\frac{P^2}{2 m \omega \hbar }} H_n\left(\frac{P}{\sqrt{m \omega \hbar} }\right),
\end{equation}
where $\mathcal{N}$ is a normalization to be set considering that
\begin{equation}
  \int_{P_{\text{min}}}^{P_{\text{max}}} |\phi_n(P)|^2 ~ \diff P = 1,
\end{equation}
with ${P_{\text{min}}}$ and ${P_{\text{max}}}$ two two bounds on $\Ph$.

To find the energy, let us start from the classical equations of motion.
Since the Hamiltonian \eqref{eqn:Hamiltonian_HO}, written in terms of canonical variables, is identical to the ordinary one, we have
\begin{align}
  q(t) &= \sqrt{\frac{2 H}{m \omega^2}} \cos(\omega t), \\
  P(t) &= - \sqrt{2 m H} \sin(\omega t)
  = \mp \sqrt{2 m H} \sqrt{1 - \frac{m \omega^2 q^2(t)}{2 H}},
  \label{eqn:HO_P}
\end{align}
where we have expressed the two quantities in terms of the constant value of the classical Hamiltonian $H$.
The two signs in the last expression correspond to different directions of oscillation.
Specifically, the negative sign corresponds to a movement to the left, while the positive sign a movement to the right.

\begin{figure}[t]
  \centering
  \includegraphics[width=\columnwidth]{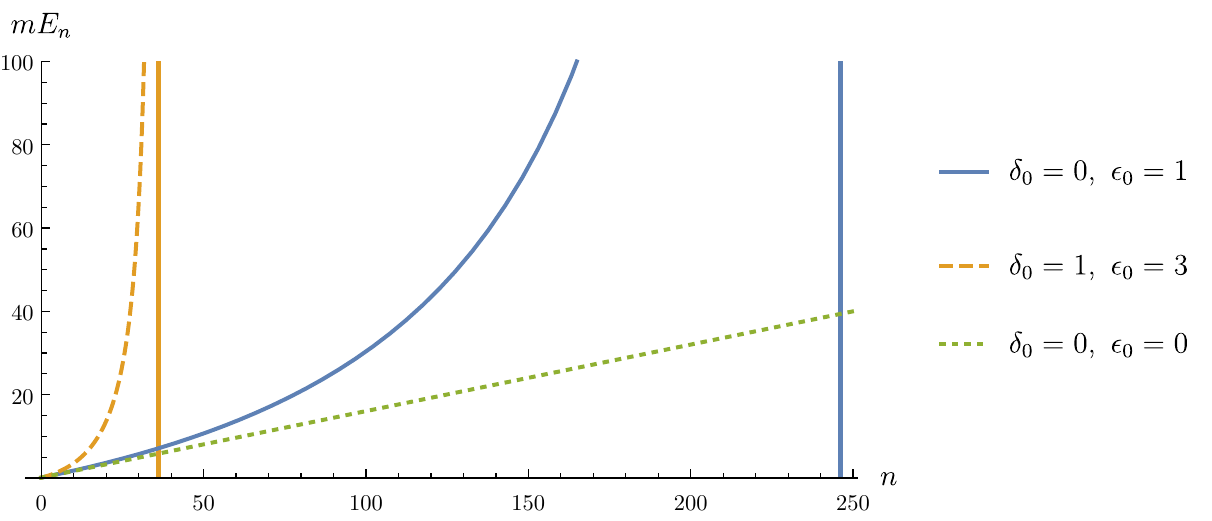}
  \caption{Energy spectrum $E_n$ of a harmonic oscillator as a function of $n$ for three different models:
    \mbox{$\delta_0=0$, $\epsilon_0=1$} \cite{Kempf:1994su} (solid, blue line);
    $\delta_0=1$, $\epsilon_0=3$ \cite{Ali:2011fa} (dashed, orange line);
    ordinary quantum mechanics (dotted, green line).
    Furthermore, the vertical lines correspond to the maximum value of $n$ allowed in the given models.
    We used a value $p_0 = 0.05 p_\text{P}$ for the characteristic momentum of the oscillator.
    The vertical axis is given in SI units.}
  \label{fig:energy_spectrum}
\end{figure}

Using \eqref{eqn:potential_energy}, we can find the potential energy for a harmonic oscillator, obtaining, up to an arbitrary constant,
\begin{multline}
  U_E
  = \frac{\sqrt{\epsilon} P(q)}{m (\delta^2 + \epsilon)} \frac{\delta \cos \left(\sqrt{\epsilon} P(q)\right)-\sqrt{\epsilon } \sin \left(\sqrt{\epsilon} P(q)\right)}{\delta \sin \left(\sqrt{\epsilon} P(q)\right) + \sqrt{\epsilon} \cos \left(\sqrt{\epsilon} P(q)\right)}\\
  -\frac{\log \left[\sqrt{\epsilon} \cos \left(\sqrt{\epsilon} P(q)\right)+\delta \sin \left(\sqrt{\epsilon} P(q)\right)\right]}{m \left(\delta ^2+\epsilon \right)},
  \label{eqn:HO_potential_energy}
\end{multline}
where, for compactness, we have used used the notiation $P(q)$ to indicate the expression of $P$ as a function of $q$ as in \eqref{eqn:HO_P}.
It is worth observing that, as a classical quantity function of $P$, this expression is equal to the opposite of the kinetic energy in \eqref{eqn:ke_2degree}, thus confirming the conservation of energy.
As for the arbitrary constant, it can be set so that the lowest value of the potential energy in a period vanishes.
Then, since the Hamiltonian can be written in terms of the momentum oscillation amplitude $\tilde{P}$ as $H = \frac{\tilde{P}^2}{2 m}$, while the total energy is only kinetic when the potential energy is zero, we have for the total energy
\begin{multline}
  E 
  = \frac{1}{m \left(\delta^2+\epsilon \right)} \left\{ \frac{\sqrt{2 H m \epsilon} \left[\frac{|\delta|}{\sqrt{\epsilon}} + \tan \left(\sqrt{2 H m \epsilon}\right)\right]}{1-\frac{|\delta|}{\sqrt{\epsilon }} \tan \left(\sqrt{2 H m \epsilon }\right)} \right. \\
  \left. + \log \left[\cos \left(\sqrt{2 H m \epsilon }\right) -\frac{|\delta|}{\sqrt{\epsilon }} \sin \left(\sqrt{2 H m \epsilon }\right)\right] \right\}.
  \label{eqn:Energy_Hamiltonian}
\end{multline}
Promoting all physical quantities to operators, we then find that the energy operator $\hat{E}$ and the Hamiltonian $\hat{H}$ are one function of the other.
Thus they commute when considered on the intersection of the respective domains.
On such an intersection, they share the same set of eigenstates.
However, the energy and Hamiltonian eigenvalues are different.
Specifically, by \eqref{eqn:Energy_Hamiltonian}, the energy eigenvalue associated to the eigenstate $|n\rangle$ in the intersection of the domains is
\begin{multline}
  E_n
  = \frac{1}{m \left(\delta^2+\epsilon \right)} \\
  \times \left\{ \frac{p_0 \sqrt{2 \epsilon (2 n + 1)} \left[\frac{|\delta|}{\sqrt{\epsilon}} + \tan \left(p_0 \sqrt{2 \epsilon (2 n + 1)}\right)\right]}{1-\frac{|\delta|}{\sqrt{\epsilon }} \tan \left(p_0 \sqrt{2 \epsilon (2 n + 1)}\right)} \right. \\
  + \log \left[\cos \left(p_0 \sqrt{2 \epsilon (2 n + 1)}\right) \right.\\
  \left. \left. -\frac{|\delta|}{\sqrt{\epsilon}} \sin \left(p_0 \sqrt{2 \epsilon (2 n + 1)}\right)\right] \right\}.
\end{multline}
It is worth noticing that the spectrum diverges when $n \to n_{\text{max}}$, as show in Figure \ref{fig:energy_spectrum}.
Thus, the bound of the Hamiltonian is associated with a diverging energy and the two operators $\hat{H}$ and $\hat{E}$ share the same domain.
In other words, energy eigenstates are stationary states and viceversa.

\section{Three spatial dimensions}
\label{sec:3d}

Let us now turn to a generic three-dimensional model of the form
\begin{equation}
  [\qh^i,\ph_j]
  = i \hbar \left[ \delta^i_j f(\ph^2) + \frac{\ph^i \ph_j}{\ph^2} g(\ph^2) \right]
  = i \hbar ~\bar{f}^i_j (\vec{\ph}),
  \label{eqn:GUP_3}
\end{equation}
with $\ph^2 = \ph^i \ph_i$.
Similar considerations presented for the one-dimensional case apply here as well, the main difference being the presence of rotations as further space symmetries.
Thus, in this case, a generic transformation consists of a displacement characterized by a vector $\vec{d}$, a time translation by an interval $\tau$, a rotation $R_{\mathbf{n}}(\theta)$ about some generic axis $\mathbf{n}$ and by an angle $\theta$, and a change of reference frame with relative velocity $\vec{v}$
\begin{align}
  \vec{q} \to & \vec{q}' = R_{\mathbf{n}} \vec{q} + \vec{d} + \vec{v} t,&
  t \to & t' = t + \tau.
  \label{eqns:transformations_3d}
\end{align}
Considering the three-dimensional extension of the operators in (\ref{eqn:time_evol_gen}-\ref{eqn:change_ref_frame_gen}), with the addition of the generator of rotations
\begin{align}
&& \centering \text{Unitar}&\text{y operator} & \centering \text{Gener}&\text{ator}\nonumber\\
  &\text{Rotation about $\mathbf{n}$}&
  \hat{U}_{\mathbf{n}} (\theta) &= e^{- i \mathbf{n} \cdot \hat{\vec{J}} \theta}&
  \hat{\vec{J}},
  \label{eqn:rotation_gen}
\end{align}
the transformation laws in \eqref{eqns:transformations_3d} imply the following set of commutation relations \cite{ballentine2014quantum}
\begin{subequations}
  \begin{align}
    [\hat{Q}^i,\hat{Q}^j] &= 0,&
    [\Ph_i,\Ph_j] &= 0,
    \label{eqns:commutators_3d_1}
    \\
    [\hat{J}_i,\hat{J}_j] &= i \hbar \epsilon_{ijk} \hat{J}_k,&
    [\hat{Q}^i,\Ph_j] &= i \hbar \delta^i_j,\\
    [\hat{Q}^i,\hat{J}_j] &= i \hbar \epsilon^i{}_{jk} \hat{Q}^k,&
    [\Qh^i,\hat{H}] &= i \hbar \Ph^i /m,\\
    [\Ph_i,\hat{J}_j] &= i \hbar \epsilon_{ij}{}^k \Ph_k,&
    [\Ph_i,\hat{H}] &= 0,\\
    [\hat{J}_i,\hat{H}] &= 0.
  \end{align}
  \label{eqns:commutators_3d}
\end{subequations}
A model such as \eqref{eqn:GUP_3} implies in general non-commutative coordinates.
To see this, it is convenient to introduce the vector of operators $\vec{\hat{\slashed Q}}$ such that $[\hat{\slashed Q}^i,\ph_j] = i \hbar \delta^i_j$.
Then, it is easy to see that, up to an uninfluencial ordering prescription \cite{Bosso:2021koi},
\begin{equation}
  \qh^i = h^i_j (\vec{\ph}) \hat{\slashed Q}^j.
\end{equation}
Finally, we can write
\begin{equation}
  [\qh^i,\qh^j] = 2 i \hbar \left[ (\hat{f} + \hat{g}) \hat{f}' - \frac{\hat{f} \hat{g}}{2 \ph^2} \right] (\ph^i \hat{\slashed Q}^j - \ph^j \hat{\slashed Q}^i),
  \label{eqn:commutator_coordinates}
\end{equation}
where for ease of notation we wrote $\hat{f} = f(\ph^2)$ and $\hat{g} = g(\ph^2)$, and where the prime indicates a derivative with respect to $\ph^2$.
It is then possible to find that Galilean transformations \eqref{eqns:transformations_3d} are not compatible with non-commutative coordinates.
Specifically, considering spatial translations, as done for one spatial dimension, it is possible to prove that
\begin{equation}
  [\qh^i,\Ph_j] = i \hbar \delta^i_j.
\end{equation}
Thus, both vectors of operators $\vec{\Qh}$ and $\vec{\qh}$ fullfil similar commutation relations.
However, because of \eqref{eqns:commutators_3d_1}, we cannot identify the two vectors.
Let us then consider the difference $\vec{\qh} - \vec{\Qh}$.
Such a difference clearly commutes with $\vec{\Ph}$.
Therefore, the difference between the operators $\vec{\qh}$ and $\vec{\Qh}$ has to depend at most on $\vec{\Ph}$.
Let us thus assume that we can write
\begin{equation}
  \qh^i = \Qh^i + \frac{\Ph^i}{\Ph} W (\Ph^2),
\end{equation}
with $\Ph = \sqrt{\Ph^i \Ph_i}$ and $W$ a function of $\Ph^2$ only.
It is then easy to find that
\begin{equation}
  [\qh^i,\qh^j]
  = \left[ \Qh^i, \frac{\Ph^j}{\Ph} W(\Ph^2) \right] - \left[ \Qh^j, \frac{\Ph^i}{\Ph} W(\Ph^2) \right]
  = 0.
\end{equation}
We then obtained that Galilean transformations, imposing the set of commutators in \eqref{eqns:commutators_3d}, necessarily require commutative coordinates.
This is consistent with what considered, for example, in Doubly Special Relativity and non-commutative geometries \cite{Amelino-Camelia:2002cqb,Calmet:2004ii,Galan:2007ns}.
Thus, from here on, we will consider only the case of commutative coordinates while maintaining Galilean symmetry, consistently with the scope of this paper.
Specifically, from \eqref{eqn:commutator_coordinates}, we can find a relation between the two functions $f$ and $g$ in \eqref{eqn:GUP_3}, namely
\begin{equation}
  g(\ph^2) = \frac{2 \ph^2 \hat{f}'}{1 - 2 \ph^2 \frac{\hat{f}'}{\hat{f}}}.
\end{equation} 
In this case, we have that $\vec{\hat{Q}} = \vec{\hat{q}}$, \emph{i.e.}, the generator of changes of reference frame $\vec{\Qh}$ can be identified with the position operator $\vec{\qh}$.
Similarly, we find that $\vec{\ph} \neq \vec{\Ph}$, but
\begin{equation}
  [\qh^i,\ph_j]
  = i \hbar \widehat{\frac{\partial p_j}{\partial P_i}} \qquad \Rightarrow \qquad
  \widehat{\frac{\partial p_j}{\partial P_i}}
  = \bar{f}^i_j (\ph^2).
\end{equation}
As for the generator of rotations, we find
\begin{equation}
  \vec{\hat{J}} = \vec{\Qh} \times \vec{\Ph},
\end{equation}
with the possible presence of a further quantity $\vec{\hat{S}}$ in case of systems with an internal structure.
We will not proceed further with the physical identification of the quantity associated with $\vec{\hat{J}}$ as this is not the scope of the present work.
Rather, we notice that, in case of a free, spinless system, and following the same arguments used for one spatial dimension, the generator of time evolution is
\begin{equation}
  \hat{H}
  = \frac{\Ph^2}{2 m}.
\end{equation}
Once again, we find that the Hamiltonian is not the energy as the latter, following the same procedure of the one-dimensional case, results to be, for a free system,
\begin{equation}
  \hat{E} = \frac{1}{m} \int \vec{\Ph} \cdot \diff \vec{\ph}.
\end{equation}
Furthermore, we notice again that the kinetic energy is modified and, specifically, it is not equal to $\frac{\ph^2}{2m}$ nor to $\frac{\Ph^2}{2m}$.
Finally, when interactions are considered, we see that the generator of time evolution may include terms in both the kinetic and potential parts, that is
\begin{equation}
  \hat{H}
  = \frac{(\Ph^i - C^i(\vec{\hat{q}}))(\Ph_i - C_i(\vec{\qh}))}{2 m} + U(\vec{\qh}),
\end{equation}
thus indicating that a minimal coupling would affect the generator of spatial translations $\vec{\Ph}$ rather than the physical momentum $\vec{\ph}$, consistently with what presented in \cite{Bosso:2018uus}.
Thus, the total energy would be written as
\begin{equation}
  \hat{E}
  = \frac{1}{m} \int (\vec{\Ph} - \vec{C}(\vec{\ph})) \cdot \diff \vec{\ph} + U_E(\vec{\qh}).
\end{equation}
As for the potential energy $U_E$, following the same argument applied to the one-dimensional case, we find
\begin{equation}
  \diff U_E(\vec{q})
  = \bar{f}^j_i(\vec{p}(\vec{q})) \frac{\diff U}{\diff q^j} \diff q^i.
\end{equation}
where $\vec{p}(\vec{q})$ corresponds to the relation betwen the physical momentum $\vec{p}$ and the physical position $\vec{q}$ as obtained from the classical equations of motion
\begin{align}
  \dot{q}^i &= \frac{\partial H}{\partial P_i},\\
  \dot{P}_i &= - \frac{\partial H}{\partial q^i},\\
  \dot{p}_i &= \frac{\partial p_i}{\partial P_j} \dot{P}_j
  = \bar{f}^j_i \dot{P}_j.
\end{align}

Thus, similarly to what we have found for the one-dimensional case, we notice that the equations describing the motion of systems in models with a minimal length, that is, the equations involving the generator $\hat{H}$, remain identical to those of standard quantum mechanics since they are written in the ordinary form in terms of operators whose commutator is given by $[\qh^i,\Ph_j] = i \hbar \delta^i_j$.
Nonetheless, one needs to notice that the representation and features of such operators may differ from those in the ordinary case.
For example, as already mentioned above, $\vec{\Ph}$ may be bounded \cite{Bosso:2020aqm}.
As for the energy, not only such an operator is not in general the generator of time evolution, but the kinetic term does not even take the usually assumed form $\frac{\ph^2}{2m}$, thus presenting further modifications than those commonly presented in the literature.

\section{Conclusions}
\label{sec:conclusions}

The presence of a minimal uncertainty in position has become an important feature suggested by many approaches to quantum gravity.
A phenomenological proposal that would realize such a minimal uncertainty in quantum mechanics consists in deforming the position-momentum commutator, thus obtaining a modified uncertainty relation via the Robertson-Schr\"odinger relation.
However, little is usually said regarding other aspects, such as the form of the Hamiltonian and energy operators in these models, as well as the influence of space and time transformations.
In the present work, we have clarified these aspects.
To do so, we explicitly employed Galilei Relativity.
We then obtained that the generator of spatial translations, $\Ph$, corresponds to the momentum conjugate to position, in general, distinct from the physical momentum, $\ph$.
Furthermore, the generator of changes of reference frame can be identified with the position operator, $\qh$, with no room for non-commutativity of coordinates.
As a consequence, the Hamiltonian, corresponding to the generator of time evolution, retains the usual form from quantum mechanics when written in terms of $\Ph$ and $\qh$.
Therefore, the ordinary dispersion relation is recovered.
Moreover, since such a dispersion relation consists in an equation between the wavelength and frequency associated with a system, a minimal wavelength due to the presence of a minimal length implies a maximum frequency, which can be interpreted as a minimal uncertainty on measurements of time intervals.
Finally, it is worth pointing out that, depending on the specific deformation, the generator $\Ph$ may be bounded, inducing a similar feature for the Hamiltonian.

Using classical definitions and arguments, we have shown that energy and Hamiltonian are two distinct physical quantities, although related.
Specifically, we have identified kinetic and potential energy.
The kinetic energy cannot be written in terms of the physical momentum as $T = \frac{p^2}{2m}$, as usually assumed in the literature.
As for the potential energy, it is related to the potential term in the Hamiltonian with a further factor deriving from the deformation of the position-momentum commutator.
Nonetheless, given the definition of kinetic and potential energies, the total energy is still conserved when all forces can be related to a potential term in the Hamiltonian.
From a quantum point of view, the conservation of energy implies that energy eigenstates are still stationary states as energy and Hamiltonian commute.
However, the two operators have distinct spectra.

In this work, we have applied these results to three exemplificative cases, namely the problems of a particle in a box, a square potential barrier, and a harmonic oscillator.
While the first two problems result to be straightforward, with the only difference from previous works concerning a slight deformation of the energy spectrum, the most interesting features have been found for the harmonic oscillator.
Specifically, we have found that the algebra of ladder operators is not affected by the deformation of the position-momentum commutator.
However, since the Hamiltonian is bounded, only a finite number of stationary states are physical, with the possibility of having no state at all when the characteristic momentum of the oscillator is larger than the bounds on the spectrum of the canonical momentum.
Furthermore, the energy diverges when approaching the limiting stationary state.
Energy is not defined beyond such a state.

Concluding, it is worth mentioning that the results presented in this work depend on Galilean invariance.
In case other symmetries and transformations are considered, a different structure is expected to appear.

\section*{Acknowledgement}

The author thanks F. Wagner and L. Petruzziello for insightful discussions and comments.


\end{document}